\documentclass[a4paper,12pt]{article}

\usepackage{ifpdf}

\newif\ifpdf
\ifx\pdfoutput\undefined
  \pdffalse
\else
  \pdfoutput=1
  \pdftrue
\fi

\RequirePackage{xspace} %
\RequirePackage{subfigure} %
\RequirePackage[centertags]{amsmath} %
\RequirePackage{amssymb}
\RequirePackage{wrapfig} %
\RequirePackage{calc} %
\RequirePackage{ifthen}
\RequirePackage{tabularx} %
\RequirePackage{flafter} %
\RequirePackage{fancyhdr} %

\ifpdf
  \RequirePackage[pdftex]{color}%
  \RequirePackage{colortbl}%
  \RequirePackage{array}%
  \RequirePackage[pdftex]{graphicx}

  \RequirePackage[ pdftex, plainpages = false, pdfpagelabels,
                 pdfpagelayout = useoutlines,
                 bookmarks,
                 breaklinks = true,
                 linktocpage,
                 pagebackref,                      
                 colorlinks = true,
                 linkcolor = blue,
                 urlcolor  = blue,
                 citecolor = blue,
                 anchorcolor = blue,
                 hyperindex = true,
                 hyperfigures
                 ]{hyperref}

\else
  \RequirePackage{color}
  \RequirePackage{colortbl}
   \RequirePackage{array}
  \RequirePackage[dvips]{graphicx}
  \RequirePackage{hyperref}
  \usepackage{rotating}
\fi

\usepackage{makeidx} 
\usepackage{setspace} 
\usepackage{rotating} 
\usepackage{ecltree}
\usepackage{epic}
\usepackage{supertabular}  
\usepackage{color}
\usepackage{exscale}
\usepackage{fontenc}
\usepackage{ifthen}
\usepackage{latexsym}
\usepackage{makeidx}
\usepackage{syntonly}
\usepackage{inputenc}
\usepackage{graphicx}
\usepackage{setspace}
\usepackage{caption2}
\usepackage[english]{babel}
\usepackage[square, comma,numbers,sort&compress]{natbib}
\usepackage{hypernat}
\usepackage{boxedminipage}
\usepackage{framed}
\usepackage{longtable}
\usepackage[all]{hypcap}    
\usepackage{algorithm2e}
\usepackage{algorithmic}
\usepackage{lscape}
\usepackage{pdflscape}

\newcommand{\note}[1]{\marginpar[left]{\singlespace \tiny #1}}

\renewcommand{\sectionmark}[1]%
      {\markright{\thesection\ #1}} 

\renewcommand{\note}[1]{}

\doublespace 

\title
{ %
\vspace*{3.0cm} \LARGE{\bf Flow of Navier-Stokes Fluids in Cylindrical Elastic Tubes} \vspace*{4.0cm} \\
}

\author{Taha Sochi\footnote{Imaging Sciences \& Biomedical Engineering, King's College London, The Rayne
Institute, St Thomas' Hospital, London, SE1 7EH, UK. Email: taha.sochi@kcl.ac.uk.} \vspace*{5.0cm}}

\begin{document}

\maketitle %
\pagenumbering{arabic}

\newpage
\phantomsection \addcontentsline{toc}{section}{Contents} %
\tableofcontents

%

\newpage
\phantomsection \addcontentsline{toc}{section}{Abstract} \noindent
{\noindent \LARGE \bf Abstract} \vspace{0.5cm}\\
\noindent %

Analytical expressions correlating the volumetric flow rate to the inlet and outlet pressures are
derived for the time-independent flow of Newtonian fluids in cylindrically-shaped elastic tubes
using a one-dimensional Navier-Stokes flow model with two pressure-area constitutive relations.
These expressions for elastic tubes are the equivalent of Poiseuille and Poiseuille-type
expressions for rigid tubes which were previously derived for the flow of Newtonian and
non-Newtonian fluids under various flow conditions. Formulae and procedures for identifying the
pressure field and tube geometric profile are also presented. The results are validated by a finite
element method implementation. Sensible trends in the analytical and numerical results are observed
and documented.

Keywords: fluid mechanics; Navier-Stokes; one-dimensional flow; Newtonian fluids; cylindrical
elastic tubes; finite element; time-independent; blood flow.

\pagestyle{headings} %
\addtolength{\headheight}{+1.6pt}
\lhead[{Chapter \thechapter \thepage}]%
      {{\bfseries\rightmark}}
\rhead[{\bfseries\leftmark}]%
     {{\bfseries\thepage}} 
\headsep = 1.0cm               

\clearpage
\section{Introduction} \label{Introduction}

Considerable amount of work has been done in the past on the flow in rigid tubes with different
types of geometry for both Newtonian and non-Newtonian fluids using various derivation methods (see
for example \cite{Skellandbook1967, BirdbookAH1987, CarreaubookKC1997, Sochithesis2007, SochiB2008,
SochiVE2009, SochiPower2011, SochiNavier2013}). However, relatively little work has been done on
the flow in elastic tubes especially on developing closed-form analytical relations. These
relations are useful in many scientific, industrial and medical applications; an obvious example is
the flow of blood in large vessels. Most of the reported work in the literature on the flow in
elastic tubes is based on the use of numerical methods such as finite element (see for instance
\cite{FormaggiaLQ2003, SherwinFPP2003}) mainly due to the fact that since the flow in networks of
elastic tubes was the main focus of these studies numerical methods were more appropriate to use.

In the current paper, explicit analytical relations linking the volumetric flow rate to the
pressure at the inlet and outlet are derived from a one-dimensional form of the Navier-Stokes
equations for cylindrically-shaped elastic tubes with constant cross sectional area using two
pressure-area constitutive models. The flow rate formulae are validated by a finite element
implementation based on a Galerkin method with Lagrange polynomial interpolation and Gauss
quadrature integration schemes. Formulae implicitly defining the tube profile and pressure field at
each point along the tube axis are also provided, demonstrated and validated. The results presented
in this paper are especially useful in biological studies such as modeling blood flow in arteries
and veins.

\section{One-Dimensional Navier-Stokes Flow Model}

The widely used one-dimensional Navier-Stokes model describing the flow of Newtonian fluids, which
is mainly formulated to model the flow in elastic tubes, is given by the following mass and
momentum conservation principles

\begin{eqnarray}
\frac{\partial A}{\partial t}+\frac{\partial Q}{\partial z}&=&0\,\,\,\,\,\,\,\,\,\,\,\,\,
t\ge0,\,\,\, z\in[0,L]     \label{ConEq1} \\
\frac{\partial Q}{\partial t}+\frac{\partial}{\partial z}\left(\frac{\alpha
Q^{2}}{A}\right)+\frac{A}{\rho}\frac{\partial p}{\partial
z}+\kappa\frac{Q}{A}&=&0\,\,\,\,\,\,\,\,\,\,\,\,\, t\ge0,\,\,\, z\in[0,L]     \label{MomEq1}
\end{eqnarray}
where $A$ stands for the tube cross sectional area, $t$ for time, $Q$ for the volumetric flow rate,
$z$ for the space coordinate along the tube axis, $L$ for the length of tube, $\alpha$
($=\frac{\int u^{2}dA}{A\overline{u}^{2}}$ with $u$ and $\overline{u}$ being the fluid local and
mean axial speed at the tube cross section respectively) for the axial momentum flux correction
factor, $\rho$ for the fluid mass density, $p$ for the $z$-dependent pressure, and $\kappa$ for the
viscosity friction coefficient which is usually given by $\kappa = 2\pi\alpha\nu/(\alpha-1)$ with
$\nu$ being the fluid kinematic viscosity defined as the ratio of the dynamic viscosity $\mu$ to
the mass density \cite{SmithPH2002, FormaggiaLQ2003, SherwinFPP2003, RuanCZC2003,
SochiTechnical1D2013}.

In using this model we assume a laminar, axi-symmetric, Newtonian, incompressible, fully-developed
flow with insignificant gravitational forces and no-slip-at-wall boundary conditions
\cite{SochiTechnical1D2013, SochiSlip2011}. For our current purpose, we also assume a steady
time-independent flow and hence we drop the time terms in the Navier-Stokes equations. In the
following section, we use this one-dimensional Navier-Stokes formulation to derive $Q$-$p$
relations for cylindrical elastic tubes using two pressure-area constitutive relations.

\section{Deriving $Q$-$p$ Relations}

For time independent flow, the Navier-Stokes system given by Equations \ref{ConEq1} and
\ref{MomEq1}, becomes

\begin{eqnarray}
\frac{\partial Q}{\partial z}&=&0\,\,\,\,\,\,\,\,\,\,\,\,\,
z\in[0,L]     \label{ConEq2} \\
\frac{\partial}{\partial z}\left(\frac{\alpha Q^{2}}{A}\right)+\frac{A}{\rho}\frac{\partial
p}{\partial z}+\kappa\frac{Q}{A}&=&0\,\,\,\,\,\,\,\,\,\,\,\,\, z\in[0,L] \label{MomEq2}
\end{eqnarray}

The first of these equations states that $Q$ as a function of $z$ is constant. With regard to the
second equation we have

\begin{equation}
\frac{A}{\rho}\frac{\partial p}{\partial z}=\frac{A}{\rho}\frac{\partial p}{\partial
A}\frac{\partial A}{\partial z}=\frac{\partial}{\partial z}\int\frac{A}{\rho}\frac{\partial
p}{\partial A}\frac{\partial A}{\partial z}\partial z=\frac{\partial}{\partial
z}\int\frac{A}{\rho}\frac{\partial p}{\partial A}dA
\end{equation}

Hence Equation \ref{MomEq2} becomes

\begin{equation}
\frac{\partial}{\partial z}\left(\frac{\alpha Q^{2}}{A}+\int\frac{A}{\rho}\frac{\partial
p}{\partial A}dA\right)+\kappa\frac{Q}{A}=0\,\,\,\,\,\,\,\,\,\,\,\,\,
z\in\left[0,L\right]\label{genMomEq}
\end{equation}

\subsection{First $p$-$A$ Model}

For this $p$-$A$ model we assume a linear pressure-area constitutive relation and hence the
pressure is proportional to the change in cross sectional area relative to the reference area, that
is

\begin{equation}
p=\gamma\left(A-A_{o}\right)\label{pAEq1}
\end{equation}
where $p$ is the actual pressure as opposed to the reference pressure to which the reference area
is defined, $\gamma$ is the proportionality coefficient which correlates to the tube stiffness, $A$
is the tube cross sectional area at pressure $p$, and $A_{o}$ is the reference area as identified
by the reference pressure which, in this equation, is set to zero for convenience without affecting
the generality of the results. From Equation \ref{pAEq1}, we get $\frac{\partial p}{\partial
A}=\gamma$ and therefore

\begin{equation}
\int\frac{A}{\rho}\frac{\partial p}{\partial A}dA=\int\frac{A}{\rho}\gamma dA=\frac{\gamma
A^{2}}{2\rho}\end{equation}
where the constant of integration is neglected because it will eventually vanish by the action of
$z$ partial derivative operator in Equation \ref{genMomEq}. Hence Equation \ref{genMomEq} becomes

\begin{equation}
\frac{\partial}{\partial z}\left(\frac{\alpha Q^{2}}{A}+\frac{\gamma
A^{2}}{2\rho}\right)+\kappa\frac{Q}{A}=0
\end{equation}
that is

\begin{equation}
\frac{\partial}{\partial A}\left(\frac{\alpha Q^{2}}{A}+\frac{\gamma
A^{2}}{2\rho}\right)\frac{\partial A}{\partial z}+\kappa\frac{Q}{A}=0
\end{equation}

\begin{equation}
\left(-\frac{\alpha Q^{2}}{A^{2}}+\frac{\gamma A}{\rho}\right)\frac{\partial A}{\partial
z}+\kappa\frac{Q}{A}=0
\end{equation}


i.e.


\begin{equation}
\frac{\partial z}{\partial A}=\frac{\alpha Q}{\kappa A}-\frac{\gamma A^{2}}{\kappa\rho Q}
\end{equation}

On integrating both sides of this equation with respect to $A$, which is always greater than zero,
we get

\begin{equation}
z=\frac{\alpha Q}{\kappa}\ln A-\frac{\gamma}{3\kappa\rho Q}A^{3}+C
\end{equation}
where $C$ is the constant of integration which can be obtained from one of the two boundary
conditions, e.g. the inlet boundary condition where $A=A_{in}$ at $z=0$ with $A_{in}$ being the
tube inlet area, that is

\begin{equation}
C=-\frac{\alpha Q}{\kappa}\ln A_{in}+\frac{\gamma}{3\kappa\rho Q}A_{in}^{3}
\end{equation}

Hence


\begin{equation}\label{zQAEq1}
z=\frac{\alpha Q}{\kappa}\ln\left(\frac{A}{A_{in}}\right)+\frac{\gamma}{3\kappa\rho
Q}\left(A_{in}^{3}-A^{3}\right)
\end{equation}

Now, from this equation combined with the other boundary condition which defines the pressure at
the outlet, that is $A=A_{ou}$ at $z=L$ where $A_{ou}$ is the tube outlet area and $L$ is the tube
length, we obtain

\begin{equation}
L=\frac{\alpha Q}{\kappa}\ln\left(\frac{A_{ou}}{A_{in}}\right)+\frac{\gamma}{3\kappa\rho
Q}\left(A_{in}^{3}-A_{ou}^{3}\right)
\end{equation}

This equation can be transformed to a quadratic polynomial in $Q$, i.e.

\begin{equation}
\frac{\alpha}{\kappa}\ln\left(\frac{A_{ou}}{A_{in}}\right)Q^{2}-LQ+\frac{\gamma}{3\kappa\rho}\left(A_{in}^{3}-A_{ou}^{3}\right)=0
\end{equation}
with the following two roots

\begin{equation}
Q=\frac{L\pm\sqrt{L^{2}-4\frac{\alpha}{\kappa}\ln\left(A_{ou}/A_{in}\right)\frac{\gamma}{3\kappa\rho}\left(A_{in}^{3}-A_{ou}^{3}\right)}}{2\frac{\alpha}{\kappa}\ln\left(A_{ou}/A_{in}\right)}
\end{equation}

For $A_{in}>A_{ou}$, which can always be satisfied by proper labeling, the two roots are
necessarily real. For a physically viable flow consistent in direction with the pressure gradient
the negative sign should be taken to obtain a positive flow and hence

\begin{equation}\label{QElastic1}
Q=\frac{L-\sqrt{L^{2}-4\frac{\alpha}{\kappa}\ln\left(A_{ou}/A_{in}\right)\frac{\gamma}{3\kappa\rho}\left(A_{in}^{3}-A_{ou}^{3}\right)}}{2\frac{\alpha}{\kappa}\ln\left(A_{ou}/A_{in}\right)}
\end{equation}

This is due to the fact that for $A_{in}>A_{ou}$ the denominator is negative and hence to obtain a
positive flow rate the numerator should be negative as well, which is the case only if the negative
sign is taken because the square root is always greater than $L$. This relation for elastic tubes
is the equivalent of the Poiseuille equation for rigid tubes. However, for elastic tubes the flow
rate is dependent not on the pressure difference but on the actual pressure at the inlet and
outlet.

\subsection{Second $p$-$A$ Model}

For the second pressure-area constitutive relation, the pressure is proportional to the radius
change with a proportionality stiffness factor scaled by the reference area, that is

\begin{equation}\label{pAEq2}
p=\frac{\beta}{A_{o}}\left(\sqrt{A}-\sqrt{A_{o}}\right)
\end{equation}
where $p$ is the pressure, $\beta$ is the tube stiffness factor, $A_{o}$ is the reference area at
the reference pressure and $A$ is the area at pressure $p$. The tube stiffness factor for the
second $p$-$A$ model is normally defined by the following relation

\begin{equation}
\beta=\frac{\sqrt{\pi}h_{o}E}{1-\varsigma^{2}}
\end{equation}
where $h_{o}$ is the tube wall thickness at reference pressure, and $E$ and $\varsigma$ are
respectively the Young's elastic modulus and Poisson's ratio of the tube wall.

From the pressure-area constitutive relation of Equation \ref{pAEq2} we obtain $\frac{\partial
p}{\partial A}=\frac{\beta}{2A_{o}\sqrt{A}}$ and therefore

\begin{equation}
\int\frac{A}{\rho}\frac{\partial p}{\partial
A}dA=\int\frac{A}{\rho}\frac{\beta}{2A_{o}\sqrt{A}}dA=\frac{\beta}{3\rho
A_{o}}A^{3/2}
\end{equation}
where the constant of integration is ignored as in the past. Hence Equation \ref{genMomEq} becomes

\begin{equation}
\frac{\partial}{\partial z}\left(\frac{\alpha Q^{2}}{A}+\frac{\beta}{3\rho
A_{o}}A^{3/2}\right)+\kappa\frac{Q}{A}=0
\end{equation}
that is

\begin{equation}
\frac{\partial}{\partial A}\left(\frac{\alpha Q^{2}}{A}+\frac{\beta}{3\rho
A_{o}}A^{3/2}\right)\frac{\partial A}{\partial z}+\kappa\frac{Q}{A}=0
\end{equation}
i.e.

\begin{equation}
\left(-\frac{\alpha Q^{2}}{A^{2}}+\frac{\beta}{2\rho A_{o}}A^{1/2}\right)\frac{\partial A}{\partial
z}+\kappa\frac{Q}{A}=0
\end{equation}

Following similar steps to those outlined in the first model, we obtain

\begin{equation}\label{zQAEq2}
z=\frac{\alpha Q^{2}\ln\left(A/A_{in}\right)-\frac{\beta}{5\rho
A_{o}}\left(A^{5/2}-A_{in}^{5/2}\right)}{\kappa Q}
\end{equation}

From the last equation associated with the second boundary condition at the outlet, i.e. $A=A_{ou}$
at $z=L$, we obtain the following expression for the volumetric flow rate

\begin{equation}
Q=\frac{-\kappa L\pm\sqrt{\kappa^{2}L^{2}-4\alpha\ln\left(A_{in}/A_{ou}\right)\frac{\beta}{5\rho
A_{o}}\left(A_{ou}^{5/2}-A_{in}^{5/2}\right)}}{2\alpha\ln\left(A_{in}/A_{ou}\right)}
\end{equation}

Both these solutions are necessarily real for $A_{in}>A_{ou}$ which can always be satisfied for
normal flow conditions by proper labeling. For a flow which is physically-consistent in direction
with the pressure gradient, the root with the plus sign should be selected, i.e.

\begin{equation}\label{QElastic2}
Q=\frac{-\kappa L+\sqrt{\kappa^{2}L^{2}-4\alpha\ln\left(A_{in}/A_{ou}\right)\frac{\beta}{5\rho
A_{o}}\left(A_{ou}^{5/2}-A_{in}^{5/2}\right)}}{2\alpha\ln\left(A_{in}/A_{ou}\right)}
\end{equation}

This, in essence, is a relation between flow rate and pressure drop, similar to the Poiseuille law
for rigid tubes, although for elastic tubes the flow rate, as given by Equation \ref{QElastic2},
does not depend on the pressure difference, as for rigid tubes, but on the actual inlet and outlet
pressure as defined by the inlet and outlet area respectively.

\section{Finite Element Formulation}

The flow formulae derived in the previous section can be validated by the finite element method
using the weak formulation. This formulation is outlined for the first and second $p$-$A$ models in
the following two subsections. More details about the finite element technicalities and the
solution scheme using Newton-Raphson iteration are given in \cite{SochiTechnical1D2013}.

\subsection{First $p$-$A$ Model}

The Navier-Stokes system, given by Equations \ref{ConEq1} and \ref{MomEq1}, can be cast in matrix
form which is more appropriate for numerical manipulation and implementation as follow

\begin{equation}\label{matrixEq1}
\frac{\partial\boldsymbol{\mathbf{U}}}{\partial t}+\frac{\partial\mathbf{\boldsymbol{F}}}{\partial
z}+\boldsymbol{\mathbf{B}}=\mathbf{0}
\end{equation}
where

\begin{equation}
\boldsymbol{\mathbf{U}}=\left[\begin{array}{c}
A\\
Q\end{array}\right]\,\,,\,\,\,\,\,\,\,\,\,\,\,\,\,\,\,\mathbf{\boldsymbol{F}}=\left[\begin{array}{c}
Q\\
\frac{\alpha Q^{2}}{A}+\frac{\gamma A^{2}}{2\rho}\end{array}\right] \,\,,\,\,\,\,\,\,\,\,\,\,\,\,\,
\textrm{and} \,\,\,\,\,\,\,\,\,\,\,\,\,\,\,
\boldsymbol{\mathbf{B}}=\left[\begin{array}{c}0\\\kappa\frac{Q}{A}\end{array}\right]
\end{equation}

On multiplying Equation \ref{matrixEq1} by weight functions and integrating over the solution
domain, $z$, the following system is obtained

\begin{equation}\label{matrixEq2}
\int_{\Omega}\frac{\partial\mathbf{U}}{\partial
t}\cdot\boldsymbol{\omega}dz+\int_{\Omega}\frac{\partial\mathbf{F}}{\partial
z}\cdot\boldsymbol{\omega}dz+\int_{\Omega}\mathbf{B}\cdot\boldsymbol{\omega}dz=\mathbf{0}
\end{equation}
where $\Omega$ is the solution domain, and $\boldsymbol{\omega}$ is a vector of arbitrary test
functions. On integrating the second term of Equation \ref{matrixEq2} by parts, the following weak
form of the preceding 1D flow system is obtained

\begin{equation}
\int_{\Omega}\frac{\partial\mathbf{U}}{\partial
t}\cdot\boldsymbol{\omega}dz-\int_{\Omega}\mathbf{F}\cdot\frac{d\boldsymbol{\omega}}{dz}dz+\int_{\Omega}\mathbf{B}\cdot\boldsymbol{\omega}dz+[\mathbf{F}\cdot\boldsymbol{\omega}]_{\partial\Omega}=\mathbf{0}
\end{equation}
where $\partial \Omega$ is the boundary of the solution domain. This weak formulation, coupled with
suitable boundary conditions, can be used as a basis for finite element implementation in
conjunction with an iterative scheme such as Newton-Raphson method. Following a solution scheme
detailed in \cite{SochiTechnical1D2013} and based on the method of characteristics
\cite{FormaggiaLQ2003, SherwinFPF2003, SherwinFPP2003, CanicK2003, PontrelliR2003,
FormaggiaLTV2006}, the eigenvalues $\lambda_{1,2}$ and left eigenvectors $\boldsymbol{L}_{1,2}$ of
the $\mathbf{H}$ matrix, which are required for obtaining the compatibility conditions on the
boundaries, are found as follow

\begin{equation}\label{detEq}
\mathrm{det}\left(\mathbf{H}-\lambda\mathbf{I}\right)=\mathrm{det}\left(\left[\begin{array}{cc}
-\lambda & 1\\
-\frac{\alpha Q^{2}}{A^{2}}+\frac{\gamma A}{\rho} & \frac{2\alpha
Q}{A}-\lambda\end{array}\right]\right)=0
\end{equation}
where $\mathbf{H}$ is the matrix of partial derivatives of $\mathbf{F}$ with respect to
$\mathbf{U}$, that is

\begin{equation}
\mathbf{H}=\frac{\partial\mathbf{F}}{\partial\mathbf{U}}=\left[\begin{array}{cc}
0 & 1\\
-\frac{\alpha Q^{2}}{A^{2}}+\frac{\gamma A}{\rho} & \frac{2\alpha
Q}{A}\end{array}\right]
\end{equation}

On solving Equation \ref{detEq} the eigenvalues are obtained

\begin{equation}
\lambda_{1,2}=\frac{\alpha
Q}{A}\pm\sqrt{\frac{Q^{2}}{A^{2}}\left(\alpha^{2}-\alpha\right)+\frac{\gamma
A}{\rho}}
\end{equation}
which are necessarily real for $\alpha\ge1$ as it is always the case, and hence the left
eigenvectors are obtained

\begin{equation}
\boldsymbol{L}_{1,2}=\left[\begin{array}{cc}
-\alpha\frac{Q}{A}\pm\sqrt{\frac{Q^{2}}{A^{2}}\left(\alpha^{2}-\alpha\right)+\frac{\gamma A}{\rho}}
& 1\end{array}\right]
\end{equation}

The compatibility conditions for the time-independent flow arising from projecting the differential
equations in the direction of the outgoing characteristic variables at the inlet and outlet are
then obtained from

\begin{equation}
\boldsymbol{L}_{1,2}\left(\mathbf{H}\frac{\partial\mathbf{U}}{\partial z}+\mathbf{B}\right)=0
\end{equation}
that is

\begin{equation}
\left[\begin{array}{cc}
-\alpha\frac{Q}{A}\pm\sqrt{\frac{Q^{2}}{A^{2}}\left(\alpha^{2}-\alpha\right)+\frac{\gamma A}{\rho}}
& 1\end{array}\right]\left[\begin{array}{c}
\frac{\partial Q}{\partial z}\\
\left(-\frac{\alpha Q^{2}}{A^{2}}+\frac{\gamma A}{\rho}\right)\frac{\partial A}{\partial
z}+\frac{2\alpha Q}{A}\frac{\partial Q}{\partial
z}+\kappa\frac{Q}{A}\end{array}\right]=0
\end{equation}

which can be simplified to

\begin{equation}
\left(-\alpha\frac{Q}{A}\pm\sqrt{\frac{Q^{2}}{A^{2}}\left(\alpha^{2}-\alpha\right)+\frac{\gamma
A}{\rho}}\right)\frac{\partial Q}{\partial z}+\left(-\frac{\alpha Q^{2}}{A^{2}}+\frac{\gamma
A}{\rho}\right)\frac{\partial A}{\partial z}+\frac{2\alpha Q}{A}\frac{\partial Q}{\partial
z}+\kappa\frac{Q}{A}=0
\end{equation}

\subsection{Second $p$-$A$ Model}

Following a similar procedure to that outlined in the previous subsection for the first $p$-$A$
model, the finite element formulation leads to the following matrix structure, eigenvalues, left
eigenvectors and time-independent compatibility conditions respectively

\begin{equation}\label{MatrixEq3}
\mathbf{U}=\left[\begin{array}{c}
A\\
Q\end{array}\right]\,\,,\,\,\,\,\,\,\,\,\,\,\,\,\,\mathbf{F}=\left[\begin{array}{c}
Q\\
\frac{\alpha Q^{2}}{A}+\frac{\beta A^{3/2}}{3\rho A_o}\end{array}\right]
\,\,\,\,\,\,\,\,\,\,\,\,\,\, \textrm{and} \,\,\,\,\,\,\,\,\,\,\,\,\,\,
\mathbf{B}=\left[\begin{array}{c}0\\\kappa\frac{Q}{A}\end{array}\right]
\end{equation}

\begin{equation}
\lambda_{1,2}=\alpha\frac{Q}{A}\pm\sqrt{\frac{Q^{2}}{A^{2}}\left(\alpha^{2}-\alpha\right)+\frac{\beta
\sqrt{A}}{2\rho A_{o}}}
\end{equation}

\begin{equation}
\boldsymbol{L}_{1,2}=\left[\begin{array}{cc}
-\alpha\frac{Q}{A}\pm\sqrt{\frac{Q^{2}}{A^{2}}\left(\alpha^{2}-\alpha\right)+\frac{\beta\sqrt{A}}{2\rho
A_{o}}} & 1\end{array}\right]\end{equation}

and

\begin{equation}
\left(-\alpha\frac{Q}{A}\pm\sqrt{\frac{Q^{2}}{A^{2}}\left(\alpha^{2}-\alpha\right)+\frac{\beta\sqrt{A}}{2\rho
A_{o}}}\right)\frac{\partial Q}{\partial z}+\left(-\alpha\frac{Q^{2}}{A^{2}}+\frac{\beta
\sqrt{A}}{2\rho A_{o}}\right)\frac{\partial A}{\partial z}+\left(2\alpha\frac{\partial Q}{\partial
z}+\kappa\right)\frac{Q}{A}=0
\end{equation}

\section{Numerical Validation}

To validate the derived flow formulae, the finite element formulation as outlined in the previous
section was implemented for the two $p$-$A$ models in a computer code using a Galerkin method with
a Lagrange polynomial interpolation associated with a Gauss quadrature integration scheme. The
comparison between the analytic and finite element solutions is outlined for some typical cases in
the following two subsections.

\subsection{First $p$-$A$ Model}

Extensive tests have been carried out to verify Equation \ref{QElastic1}; a sample of which is
given in Table \ref{FirstSampleTable}. Certain sensible trends can be observed in these results.
For example, the diagonally-oriented entries from top-left to bottom-right direction in the table
are of similar magnitude which is sensible since in this quasi-linear flow regime obtained at
relatively low pressures the flow is Poiseuille-like and hence it is almost proportional to the
pressure difference (i.e. $P_{in}-P_{ou}$). This Poiseuille-like behavior disappears at
high-pressure flow regimes as the flow rate becomes increasingly dependent on the actual pressures
at the inlet and outlet rather than on the pressure difference. Another sensible trend is that the
flow rate in these diagonally-oriented entries is increasing in the top-left to bottom-right
direction due to the fact that although the pressure difference for these entries is the same, the
lower entries have larger area at the inlet and outlet, due to the higher pressure at the tube
entrance and exit, than the upper ones. This trend is more obvious at higher pressure regimes.

\begin{figure}[!h]
\centering{}
\includegraphics
[scale=0.6] {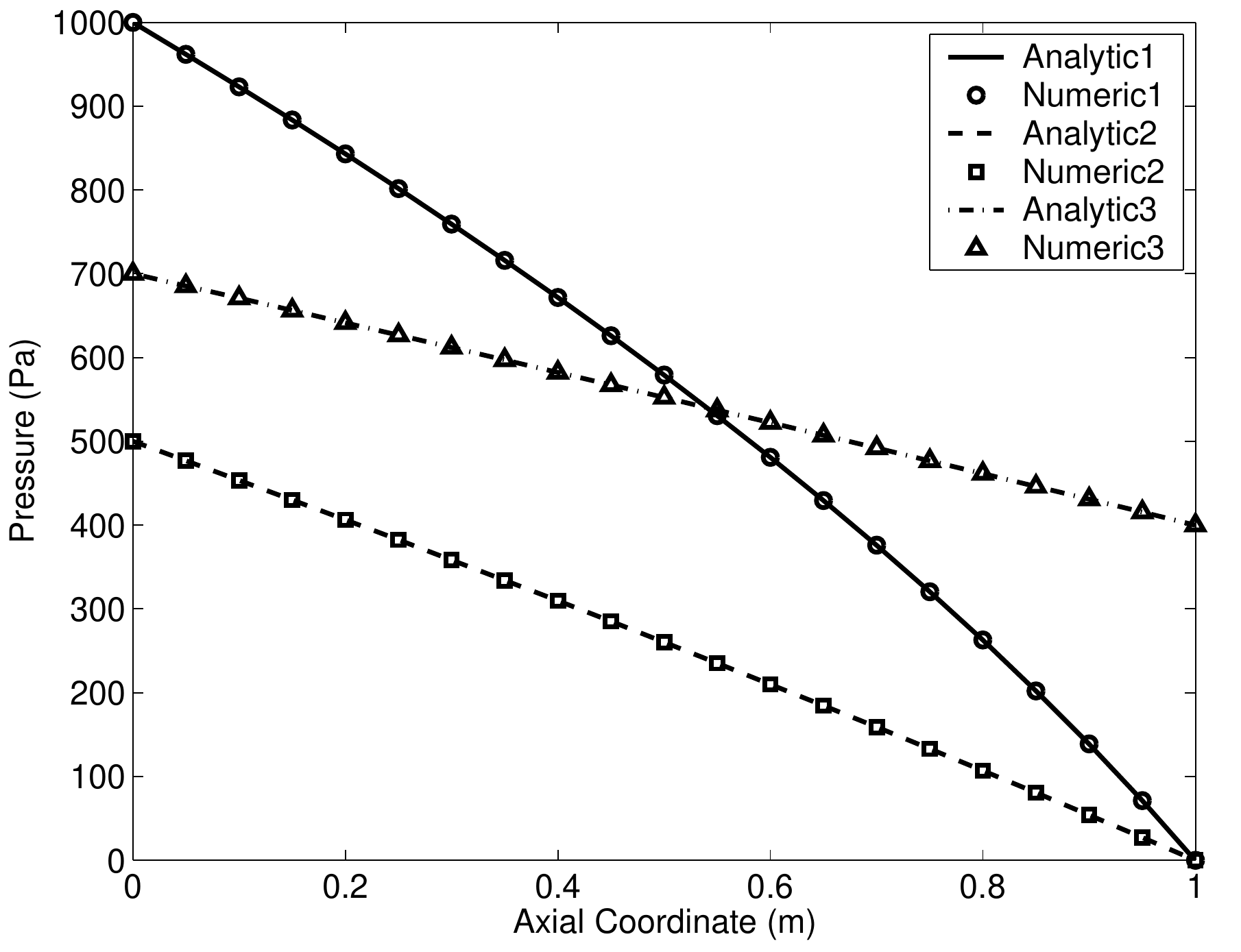} \caption{Pressure versus tube axial coordinate for three sample cases related
to the first $p$-$A$ model as obtained analytically from Equation \ref{zQAEq1} and numerically by a
finite element method with a quadratic polynomial interpolation scheme as outlined in the previous
section. The labels `1', `2' and `3' in these plots refer respectively to the cases where
$P_{in}=1000$~Pa and $P_{ou}=0$~Pa, $P_{in}=500$~Pa and $P_{ou}=0$~Pa, and $P_{in}=700$~Pa and
$P_{ou}=400$~Pa. The tube, fluid and flow parameters with which these results are obtained are:
$\rho=1060$~kg/m$^3$, $\mu=0.0035$~Pa.s, $\alpha=1.333$, $L=1.0$~m, $r=0.1$~m, and
$\gamma=5\times10^6$~Pa/m$^2$.} \label{P1Fig}
\end{figure}

\begin{figure}[!h]
\centering{}
\includegraphics
[scale=0.6] {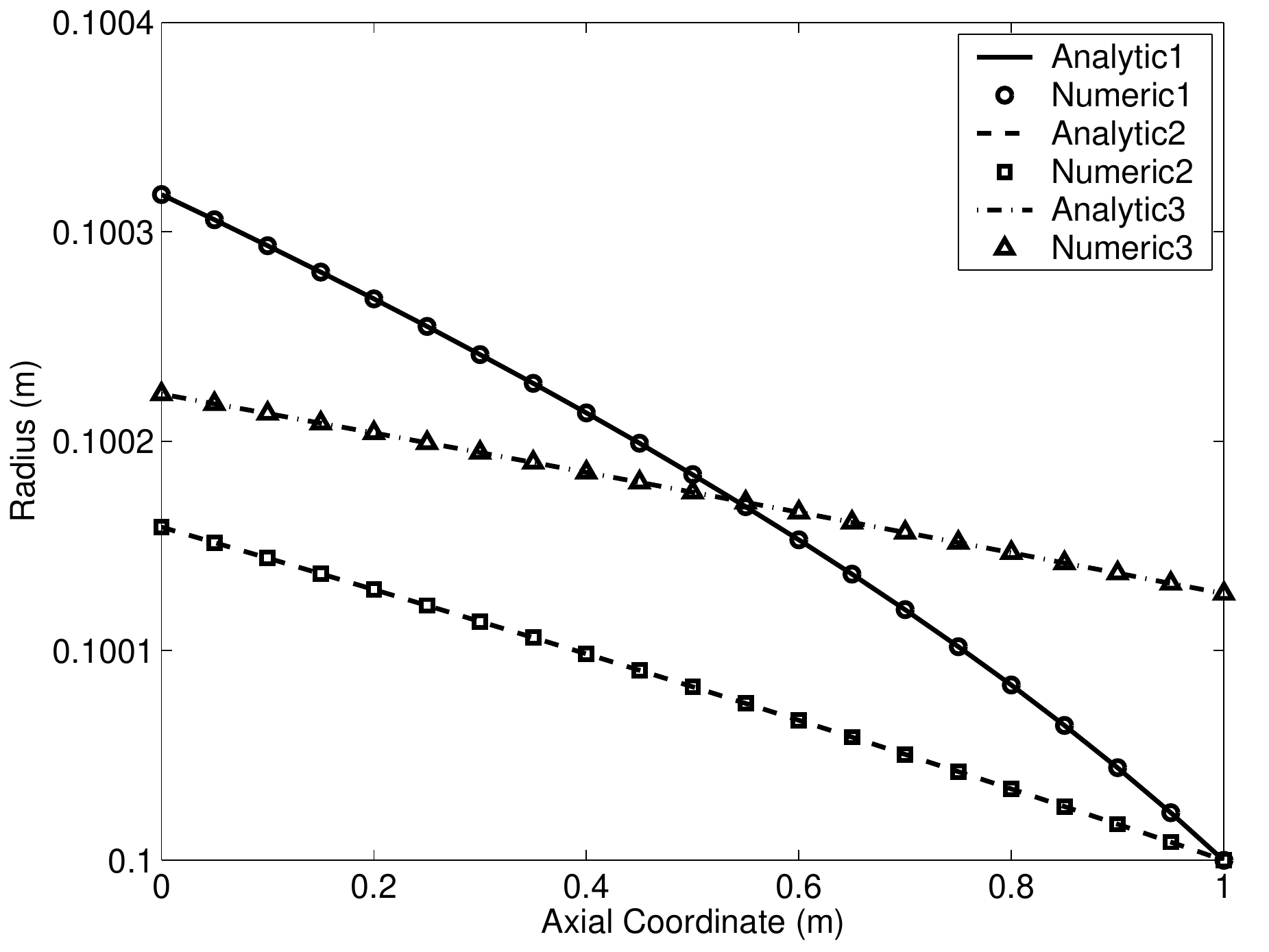} \caption{Radius versus tube axial coordinate for the three sample cases of
Figure \ref{P1Fig}.} \label{R1Fig}
\end{figure}

We also used Equation \ref{zQAEq1}, which implicitly correlates $A$ to $z$, to obtain the pressure
field inside the tube and the tube profile by numerically solving for $A$ for a given $z$. A sample
of these results, with their finite element counterparts, is presented in Figures \ref{P1Fig} and
\ref{R1Fig}. These figures confirm the sensibility of the obtained analytical and numerical
results.

\begin{table} [!h]
\caption{Sample results of the volumetric flow rate in m$^3$/s related to the elastic tube
investigation for the first $p$-$A$ model. The rows stand for the inlet pressure, $P_{in}$, and the
columns for the outlet pressure, $P_{ou}$, in Pa. The parameters with which these results are
obtained are: $\rho=1060$~kg/m$^3$, $\mu=0.0035$~Pa.s, $\alpha=1.333$, $L=1.0$~m, $r=0.1$~m, and
$\gamma=5\times10^6$~Pa/m$^2$. In each $P_{in}$ row the top and bottom entries are respectively the
analytic solution, given by Equation \ref{QElastic1}, and the finite element solution which is
obtained with a quadratic Lagrange polynomial interpolation. \label{FirstSampleTable}}
 \vspace{-0.7cm}
\begin{center}
{\tiny
\begin{tabular}{|c|cccccccccc|}
\hline
           &                                                                                          \multicolumn{ 10}{c|}{{\bf $P_{ou}$}} \\
           &    {\bf 0} &  {\bf 100} &  {\bf 200} &  {\bf 300} &  {\bf 400} &  {\bf 500} &  {\bf 600} &  {\bf 700} &  {\bf 800} &  {\bf 900} \\
{\bf $P_{in}$} &            &            &            &            &            &            &            &            &            &            \\
\hline
\multicolumn{ 1}{|c|}{{\bf 100}} &   0.286046 &            &            &            &            &            &            &            &            &            \\
\multicolumn{ 1}{|c|}{{\bf }} &   0.286046 &            &            &            &            &            &            &            &            &            \\
\hline
\multicolumn{ 1}{|c|}{{\bf 200}} &   0.307977 &   0.286332 &            &            &            &            &            &            &            &            \\
\multicolumn{ 1}{|c|}{{\bf }} &   0.307977 &   0.286332 &            &            &            &            &            &            &            &            \\
\hline
\multicolumn{ 1}{|c|}{{\bf 300}} &   0.315789 &   0.308278 &   0.286619 &            &            &            &            &            &            &            \\
\multicolumn{ 1}{|c|}{{\bf }} &   0.315789 &   0.308278 &   0.286619 &            &            &            &            &            &            &            \\
\hline
\multicolumn{ 1}{|c|}{{\bf 400}} &   0.319850 &   0.316096 &   0.308579 &   0.286905 &            &            &            &            &            &            \\
\multicolumn{ 1}{|c|}{{\bf }} &   0.319850 &   0.316096 &   0.308579 &   0.286905 &            &            &            &            &            &            \\
\hline
\multicolumn{ 1}{|c|}{{\bf 500}} &   0.322373 &   0.320158 &   0.316402 &   0.308881 &   0.287192 &            &            &            &            &            \\
\multicolumn{ 1}{|c|}{{\bf }} &   0.322373 &   0.320159 &   0.316402 &   0.308881 &   0.287192 &            &            &            &            &            \\
\hline
\multicolumn{ 1}{|c|}{{\bf 600}} &   0.324118 &   0.322684 &   0.320467 &   0.316708 &   0.309182 &   0.287479 &            &            &            &            \\
\multicolumn{ 1}{|c|}{{\bf }} &   0.324119 &   0.322684 &   0.320467 &   0.316708 &   0.309182 &   0.287479 &            &            &            &            \\
\hline
\multicolumn{ 1}{|c|}{{\bf 700}} &   0.325415 &   0.324430 &   0.322994 &   0.320776 &   0.317015 &   0.309484 &   0.287766 &            &            &            \\
\multicolumn{ 1}{|c|}{{\bf }} &   0.325415 &   0.324430 &   0.322994 &   0.320777 &   0.317015 &   0.309484 &   0.287766 &            &            &            \\
\hline
\multicolumn{ 1}{|c|}{{\bf 800}} &   0.326430 &   0.325727 &   0.324741 &   0.323305 &   0.321086 &   0.317322 &   0.309785 &   0.288053 &            &            \\
\multicolumn{ 1}{|c|}{{\bf }} &   0.326430 &   0.325728 &   0.324742 &   0.323305 &   0.321086 &   0.317322 &   0.309785 &   0.288053 &            &            \\
\hline
\multicolumn{ 1}{|c|}{{\bf 900}} &   0.327256 &   0.326743 &   0.326040 &   0.325053 &   0.323615 &   0.321395 &   0.317628 &   0.310087 &   0.288340 &            \\
\multicolumn{ 1}{|c|}{{\bf }} &   0.327257 &   0.326743 &   0.326040 &   0.325053 &   0.323616 &   0.321395 &   0.317628 &   0.310087 &   0.288340 &            \\
\hline
\multicolumn{ 1}{|c|}{{\bf 1000}} &   0.327950 &   0.327569 &   0.327056 &   0.326352 &   0.325365 &   0.323926 &   0.321704 &   0.317935 &   0.310389 &   0.288627 \\
\multicolumn{ 1}{|c|}{{\bf }} &   0.327951 &   0.327570 &   0.327056 &   0.326353 &   0.325365 &   0.323926 &   0.321704 &   0.317935 &   0.310389 &   0.288627 \\
\hline
\end{tabular}
}
\end{center}
\end{table}

\clearpage
\subsection{Second $p$-$A$ Model}

\begin{figure}[!h]
\centering{}
\includegraphics
[scale=0.6] {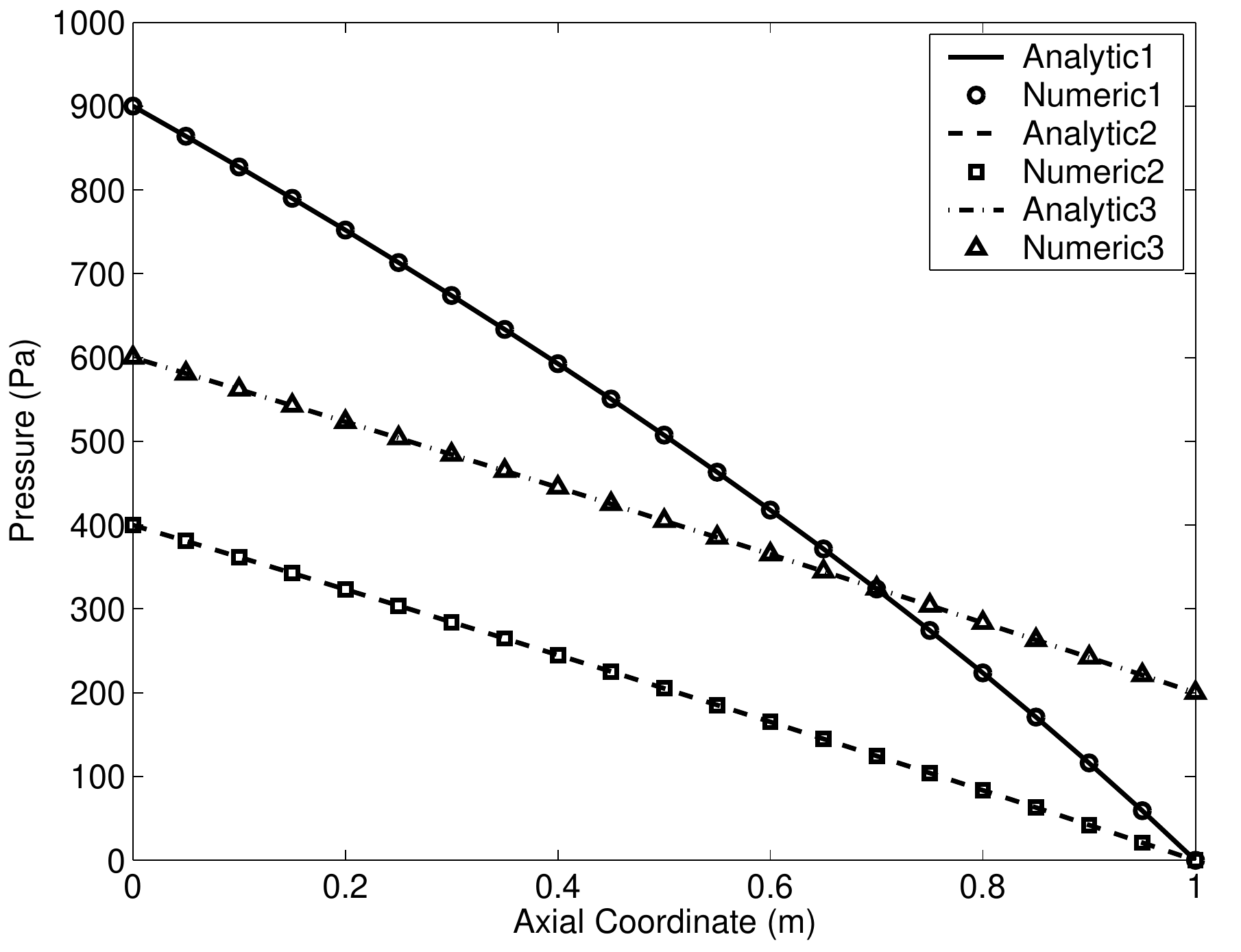} \caption{Pressure versus tube axial coordinate for three sample cases related
to the second $p$-$A$ model as obtained analytically from Equation \ref{QElastic2} and numerically
by a finite element method with a quadratic polynomial interpolation scheme as outlined in the
previous section. The labels `1', `2' and `3' in these plots refer respectively to the cases where
$P_{in}=900$~Pa and $P_{ou}=0$~Pa, $P_{in}=400$~Pa and $P_{ou}=0$~Pa, and $P_{in}=600$~Pa and
$P_{ou}=200$~Pa. The tube, fluid and flow parameters with which these results are obtained are:
$\rho=1060$~kg/m$^3$, $\mu=0.0035$~Pa.s, $\alpha=1.333$, $L=1.0$~m, $r=0.1$~m, and
$\beta=5\times10^4$~Pa.m.} \label{P2Fig}
\end{figure}

\begin{figure}[!h]
\centering{}
\includegraphics
[scale=0.6] {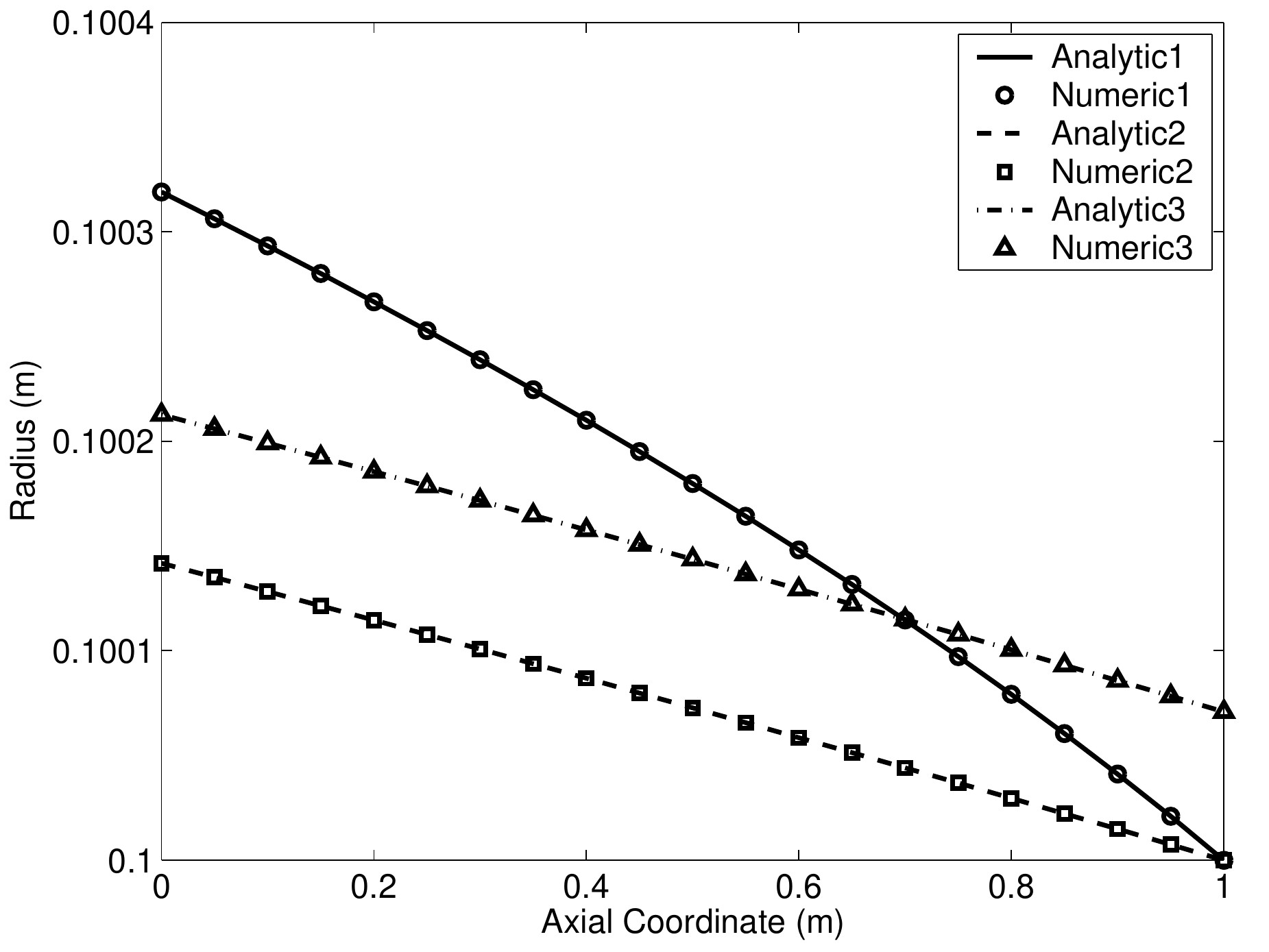} \caption{Radius versus tube axial coordinate for the three sample cases of
Figure \ref{P2Fig}.} \label{R2Fig}
\end{figure}

Extensive tests have been carried out to verify Equation \ref{QElastic2}; a sample of which is
given in Table \ref{SecondSampleTable}. Also, we used Equation \ref{zQAEq2} to obtain the pressure
field inside the tube and the tube geometric profile, as outlined for the first $p$-$A$ model. A
sample of these results, with their finite element equivalents, is presented in Figures \ref{P2Fig}
and \ref{R2Fig}. Similar sensible trends to those observed in the first $p$-$A$ model are detected.

\begin{table} [!h]
\caption{Sample results of the volumetric flow rate in m$^3$/s related to the elastic tube
investigation for the second $p$-$A$ model. The rows stand for the inlet pressure, $P_{in}$, and
the columns for the outlet pressure, $P_{ou}$, in Pa. The parameters with which these results are
obtained are: $\rho=1060$~kg/m$^3$, $\mu=0.0035$~Pa.s, $\alpha=1.333$, $L=1.0$~m, $r=0.1$~m, and
$\beta=5\times10^4$~Pa.m. In each $P_{in}$ row the top and bottom entries are respectively the
analytic solution, given by Equation \ref{QElastic2}, and the finite element solution which is
obtained with a quadratic Lagrange polynomial interpolation. \label{SecondSampleTable}}
 \vspace{-0.7cm}
\begin{center}
{\tiny
\begin{tabular}{|c|cccccccccc|}
\hline
           &                                                                                          \multicolumn{ 10}{c|}{{\bf $P_{ou}$}} \\
           &    {\bf 0} &  {\bf 100} &  {\bf 200} &  {\bf 300} &  {\bf 400} &  {\bf 500} &  {\bf 600} &  {\bf 700} &  {\bf 800} &  {\bf 900} \\
{\bf $P_{in}$} &            &            &            &            &            &            &            &            &            &            \\
\hline
\multicolumn{ 1}{|c|}{{\bf 100}} &   0.273135 &            &            &            &            &            &            &            &            &            \\
\multicolumn{ 1}{|c|}{{\bf }} &   0.273135 &            &            &            &            &            &            &            &            &            \\
\hline
\multicolumn{ 1}{|c|}{{\bf 200}} &   0.292950 &   0.273397 &            &            &            &            &            &            &            &            \\
\multicolumn{ 1}{|c|}{{\bf }} &   0.292950 &   0.273397 &            &            &            &            &            &            &            &            \\
\hline
\multicolumn{ 1}{|c|}{{\bf 300}} &   0.299986 &   0.293221 &   0.273659 &            &            &            &            &            &            &            \\
\multicolumn{ 1}{|c|}{{\bf }} &   0.299986 &   0.293221 &   0.273659 &            &            &            &            &            &            &            \\
\hline
\multicolumn{ 1}{|c|}{{\bf 400}} &   0.303637 &   0.300259 &   0.293491 &   0.273922 &            &            &            &            &            &            \\
\multicolumn{ 1}{|c|}{{\bf }} &   0.303637 &   0.300259 &   0.293491 &   0.273922 &            &            &            &            &            &            \\
\hline
\multicolumn{ 1}{|c|}{{\bf 500}} &   0.305904 &   0.303912 &   0.300532 &   0.293762 &   0.274184 &            &            &            &            &            \\
\multicolumn{ 1}{|c|}{{\bf }} &   0.305904 &   0.303912 &   0.300533 &   0.293762 &   0.274184 &            &            &            &            &            \\
\hline
\multicolumn{ 1}{|c|}{{\bf 600}} &   0.307471 &   0.306180 &   0.304187 &   0.300806 &   0.294033 &   0.274447 &            &            &            &            \\
\multicolumn{ 1}{|c|}{{\bf }} &   0.307471 &   0.306180 &   0.304187 &   0.300806 &   0.294033 &   0.274447 &            &            &            &            \\
\hline
\multicolumn{ 1}{|c|}{{\bf 700}} &   0.308634 &   0.307747 &   0.306455 &   0.304462 &   0.301080 &   0.294304 &   0.274710 &            &            &            \\
\multicolumn{ 1}{|c|}{{\bf }} &   0.308634 &   0.307747 &   0.306456 &   0.304462 &   0.301080 &   0.294304 &   0.274710 &            &            &            \\
\hline
\multicolumn{ 1}{|c|}{{\bf 800}} &   0.309543 &   0.308910 &   0.308023 &   0.306731 &   0.304737 &   0.301353 &   0.294575 &   0.274973 &            &            \\
\multicolumn{ 1}{|c|}{{\bf }} &   0.309543 &   0.308910 &   0.308023 &   0.306731 &   0.304737 &   0.301353 &   0.294575 &   0.274973 &            &            \\
\hline
\multicolumn{ 1}{|c|}{{\bf 900}} &   0.310283 &   0.309820 &   0.309187 &   0.308299 &   0.307007 &   0.305012 &   0.301627 &   0.294847 &   0.275236 &            \\
\multicolumn{ 1}{|c|}{{\bf }} &   0.310283 &   0.309820 &   0.309187 &   0.308300 &   0.307007 &   0.305012 &   0.301627 &   0.294847 &   0.275237 &            \\
\hline
\multicolumn{ 1}{|c|}{{\bf 1000}} &   0.310903 &   0.310560 &   0.310097 &   0.309464 &   0.308576 &   0.307283 &   0.305287 &   0.301902 &   0.295118 &   0.275500 \\
\multicolumn{ 1}{|c|}{{\bf }} &   0.310904 &   0.310561 &   0.310097 &   0.309464 &   0.308576 &   0.307283 &   0.305287 &   0.301902 &   0.295118 &   0.275500 \\
\hline
\end{tabular}
}
\end{center}
\end{table}

\clearpage
\section{Conclusions} \label{Conclusions}

In this study, two analytical expressions, correlating volumetric flow rate to pressure at inlet
and outlet, are derived for the Newtonian flow in cylindrical elastic tubes from a one-dimensional
form of the Navier-Stokes equations. The expressions are validated numerically by a finite element
method based on a Galerkin scheme with Lagrange interpolation and Gauss quadrature integration.
Sample results, which are quantitatively and qualitatively sensible, are presented for
demonstration. Two constitutive relations, depicting the nature of the relation between area and
pressure in elastic tubes, are used in all these derivations and finite element implementation. The
foundations of the finite element weak form for the two $p$-$A$ models are outlined for completion.
Preliminary rational trends in these results are observed and documented. Analytical implicit
relations for obtaining the pressure field inside the tube, as well as the tube geometric profile,
are also presented, demonstrated and numerically validated. The outcome of this investigation,
numerical as well as analytical, is of relevance to several areas of science, technology and
medicine.

\clearpage
\phantomsection \addcontentsline{toc}{section}{Nomenclature} %
{\noindent \LARGE \bf Nomenclature} \vspace{0.5cm}

\begin{supertabular}{ll}
$\alpha$                &   correction factor for axial momentum flux \\
$\beta$                 &   stiffness factor in the second $p$-$A$ model \\
$\gamma$                &   stiffness factor in the first $p$-$A$ model \\
$\kappa$                &   viscosity friction coefficient \\
$\lambda_{1,2}$         &   eigenvalues of $\mathbf{H}$ matrix \\
$\mu$                   &   fluid dynamic viscosity \\
$\nu$                   &   fluid kinematic viscosity \\
$\rho$                  &   fluid mass density \\
$\varsigma$             &   Poisson's ratio of tube wall \\
$\boldsymbol{\omega}$   &   vector of test functions in finite element formulation \\
$\Omega$                &   solution domain \\
$\partial \Omega$       &   boundary of solution domain \\
\\
$A$                     &   tube cross sectional area at pressure $p$ \\
$A_{in}$                &   tube cross sectional area at inlet \\
$A_o$                   &   tube reference cross sectional area at reference pressure \\
$A_{ou}$                &   tube cross sectional area at outlet \\
$\mathbf{B}$            &   matrix of force terms in the 1D Navier-Stokes equations \\
$E$                     &   Young's modulus of tube wall \\
$\mathbf{F}$            &   flux matrix in the 1D Navier-Stokes equations \\
$\mathbf{H}$            &   matrix of partial derivatives of $\mathbf{F}$ with respect to $\mathbf{U}$ \\
$h_o$                   &   tube wall thickness at reference pressure \\
$L$                     &   length of tube \\
$\boldsymbol{L}_{1,2}$  &   left eigenvectors of $\mathbf{H}$ matrix \\
$p$                     &   pressure at given coordinate $z$ \\
$P_{in}$                &   pressure at tube inlet \\
$P_{ou}$                &   pressure at tube outlet \\
$Q$                     &   volumetric flow rate \\
$r$                     &   radius \\
$t$                     &   time \\
$u$                     &   local axial speed of fluid at cross section \\
$\overline{u}$          &   mean axial speed of fluid at cross section \\
$\mathbf{U}$            &   vector of Navier-Stokes dependent variables \\
$z$                     &   tube axial coordinate \\

\end{supertabular}

\newpage
\phantomsection \addcontentsline{toc}{section}{References} %
\bibliographystyle{unsrt}

\end{document}